# ARQ for Physical-layer Network Coding

Jianghao He and Soung-Chang Liew, Fellow, IEEE

*Abstract*- **This paper investigates ARQ (Automatic Repeat request) designs for Physical-layer Network Coding (PNC) systems. Most prior work related to PNC explores its use in Two-Way Relay Channel (TWRC). We have previously found that, besides TWRC, there are many other PNC building blocks—building blocks are simple small network structures that can be used to construct a large network. In some of these PNC building blocks, the receivers can obtain side information through overhearing. Although such overheard information is not the target information that the receivers desire, the receivers can exploit the overheard information together with a network-coded packet received to obtain a desired native packet. This can yield substantial throughput gain. Our previous study, however, assumed what is sent always gets received. In practice, that is not the case. Error control is needed to ensure reliable communication. This paper focuses on ARQ designs for ensuring reliable PNC communication. The availability of overheard Information and its potential exploitation make the ARQ design of a network-coded system different from that of a non-network-coded system. In this paper, we lay out the fundamental considerations for such ARQ designs: 1) we put forth a framework to track the stored coded packets and overheard packets to increase the chance of packet extraction, and derive the throughput gain achieved therefore; 2) we investigate two variations of PNC ARQ, coupled and non-coupled ARQs, and prove that non-coupled ARQ is more efficient; 3) we show how to optimize parameters in PNC ARQ—specifically the window size and the ACK frequency—to minimize the throughput degradation caused by ACK feedback overhead and wasteful retransmissions due to lost ACK.**

*Index Terms*- Physical-layer Network Coding, Multi-hop Wireless Networks, Selective Repeat ARQ, Selective Acknowledgements, Wireless Scheduling, Overheard information, End-to-End Design, Unicast.

## I. INTRODUCTION

Physical-layer network coding (PNC) [1] can boost throughput in wireless relay networks. In a two-way relay channel (TWRC), by allowing the two end nodes to transmit simultaneously to the relay and not treating this as collision, PNC can increase the system throughput by 100% [1] compared with traditional relaying. Furthermore, as shown in [2], besides TWRC, PNC can also be applied to many other relay network structures, referred to as PNC building blocks (or PNC atoms). In essence, a large network can be decomposed into a multiplicity of PNC building blocks so that PNC can be applied in a systematic way to achieve throughput gain. Each of these building blocks can achieve throughput gain ranging from 50% to 167%, translating to an overall throughput gain of around 100% for the large network [2].

The throughput gain as demonstrated in [2] assumes transmissions in the building blocks are reliable, that is, all transmissions are successfully received. In practice, since these transmissions are over error-prone wireless links, the packets may get corrupted and not received. Automatic repeat request (ARQ) is a means for error control to ensure reliable communication. To our best knowledge, ARQ for PNC systems has not been systematically studied before. Specifically, ARQ that takes into account that PNC building blocks may exploit overheard packets to achieve throughput gain has not been studied. In this paper, we investigate the design principles of such ARQ in an attempt to answer the two sets of questions below:

   1. What is the difference between PNC ARQ and traditional ARQ? What are the design alternatives for PNC ARQ? How to improve the efficiency of the retransmissions in PNC

• *Both authors are with the Department of Information Engineering, The Chinese University of Hong Kong, Hong Kong Special Administrative Region, China. E-mail: hjh010@ ie.cuhk.edu.hk; soung@ie.cuhk.edu.hk*

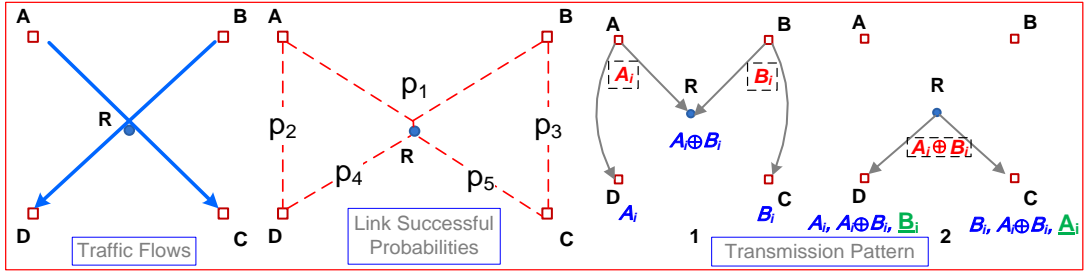

Fig. 1. PNC Cross Atom

building blocks that exploit overhearing?

2. What are the optimal throughputs of PNC building blocks with erasure channels? Can ARQ be designed to allow the PNC building blocks to achieve these optimal throughputs?

To illustrate ARQ for PNC building blocks with overhearing, let us give an example based on a specific PNC building block, referred to as the cross atom, as shown in Fig. 1. The cross atom consists of one relay, two source nodes, and two receiver nodes. There are two unicast flows, A-R-C and B-R-D. Node A wants to transmit a sequence of packets, $A_1, A_2, A_3\ldots$, to node C via relay R; node B wants to transmit a sequence of packets, $B_1, B_2, B_3\ldots$, to node D via R[1]. Suppose that 1) nodes C and D can overhear the transmissions of nodes B and A, respectively; and 2) the transmissions of nodes A and B do not affect the overhearing/reception of node B's packets by node C and the overhearing/reception of node A's packets by node D. If links are perfectly reliable, two time slots are needed to deliver one packet from A to C and one packet from B to D, as explained below. In the first timeslot, nodes A and B transmit packets $A_1$ and $B_1$, respectively. Nodes C and D overhear packets $B_1$ and $A_1$, respectively. With the PNC mechanism, node R receives packet $A_1 \oplus B_1$. In the second timeslot, node R transmits packet $A_1 \oplus B_1$, which is received by nodes C and D. Node C can then extract its target packet, packet $A_1$, by $(A_1 \oplus B_1) \oplus B_1$; similarly, node D can extract its target packet, packet $B_1$, by $(A_1 \oplus B_1) \oplus A_1$.

Now, suppose that each link is a packet erasure channel rather than a perfectly reliable channel. Suppose that in the first time slot, node C fails to overhear packet $B_1$, while node D successfully overhears packet $A_1$ and relay $R$ successfully receives $A_1 \oplus B_1$. Furthermore, in the second time slot, node D fails to receive packet $A_1 \oplus B_1$, while node C successfully receives packet $A_1 \oplus B_1$, from relay R. After one round (the two timeslots), neither node C or D can extract its desired packet (packets $A_1$ and packet $B_1$, respectively). Thus, an ARQ mechanism is needed to ensure nodes C and D can eventually obtain packets $A_1$ and packet $B_1$.

In this paper, we focus on end-to-end ARQ. For the cross atom, this means that the ARQ mechanism is between the source nodes, A and B, and the destination nodes, C and D. The relay does not take part in packet retransmission and is oblivious of the ARQ mechanism: it just relays network-coded packets.

---

[1] In this paper, we use straight font to label nodes and use the corresponding italic font to label the packets produced by the nodes.

Continuing with the example, although nodes C and D could obtain only packet $A_1 \oplus B_1$ and packet $A_1$, but not their desired packets, packet $A_1$ and packet $B_1$, respectively, the received packets are nonetheless useful in the future and should be stored. In this example, suppose we adopt an ARQ window size of one. Since nodes C and D did not receive their desired packets, no ACK will be returned to nodes A and B. In the next round, since nodes A and B did not receive ACKs, they retransmit packets $A_1$ and $B_1$, respectively, in the third timeslot (the first timeslot of the second round). Now, as long as node C can overhear packet $B_1$ in the third timeslot, it can already extract its desired packet $A_1$, since node C already has packet $A_1 \oplus B_1$ from the last round. This is the case even if node C does not receive $A_1 \oplus B_1$ from R in the fourth timeslot. By the same token, even if node D fails to overhear packet $A_1$ in the third timeslot, as long as it receives $A_1 \oplus B_1$ from R in the fourth timeslot, it can extract packet $B_1$. The storage and the reuse of overheard and coded packets can increase the chance of destination nodes extracting their desired packets in a lossy network.

With the possibility of reusing stored overheard and coded packets, an issue that does not occur in traditional ARQ arises: how should we "track" which stored packets may still be potentially useful and which stored packets will no longer be useful in the future, especially when we increase the ARQ window size to beyond one? Specifically, we need a tracking mechanism to decide which overheard packets and coded packets should continue to be stored, and which should be discarded.

This ***stored-packet tracking*** raises yet another design issue, as explained below. In general, each PNC building block has multiple unicast traffic flows between several source-destination pairs. Consider one particular flow *f*. It is natural for the source node of *f* to move on to deliver the next packet after its current packet has been delivered. For example, in the cross atom in Fig.1, after node A successfully delivers packet $A_i$ to node C, it will transmit packet $A_{i+1}$ in its next transmission. This strategy is referred to as ***non-coupled ARQ*** (its formal definition is given in Section III). It is also possible to have another strategy, referred to as ***coupled ARQ*** (formal definition also in Section III), wherein the source node of *f* may not move on to deliver its next packet even after the current packet has been successfully delivered, because the current packet of a different flow *g* has not been successfully delivered. This is the case, for example, if flow *g* wants to use the current packet of *f* to extract a desired packet. With respect to the cross PNC atom in Fig.1, for example, if nodes A and B transmit packets $A_i$ and $B_i$, and node C successfully extracts packet $A_i$ but node D fails to extract packet $B_i$. Furthermore, suppose that the reason for the failure at node D is that node D only receives packet $A_i \oplus B_i$ but misses packet $A_i$. If node A progresses to transmit packet $A_{i+1}$, then the packet $A_i \oplus B_i$ will not be useful at node D. However, if node A retransmits $A_i$, then the packet $A_i \oplus B_i$ will still be useful. In coupled ARQ, the progression of the transmissions of different flows are coupled together, hence the term "coupled ARQ". Compared with non-coupled ARQ, the tradeoff is that node A misses the chance to transmit a new packet. In this paper, we show that non-coupled ARQ is more efficient than coupled ARQ[2].

In general, not only DATA packets, ACKs may also get lost in wireless networks. Missing ACKs leads to wasteful retransmissions (i.e., packets may be retransmitted even if their target receivers already received them). Transmitting excessive numbers of ACKs to ensure their receptions, on the other hand, may lead to large feedback overhead. This paper also investigates how to set parameters in the PNC ARQ mechanism—specifically the window size and the ACK frequency (i.e., number of DATA packets per ACK)—to strike a balance to optimize throughput.

**In this paper, to answer the first set of questions as set out at the beginning of this section:**
1) We point out the importance of storing and tracking both coded packets and overheard packets, a distinguishing feature that sets PNC ARQ apart from traditional ARQ; and we investigate how to optimize the use of the stored packets to improve retransmission efficiency.
2) We show that for PNC atoms, non-coupled ARQ is more efficient than coupled ARQ, regardless of the error probabilities of wireless links.

**To answer the second set of questions as set out at the beginning of this section:**
3) We construct a Markov model to derive an idealized throughput equation that relates channel erasure probabilities with throughput. This idealized throughput is an upper bound for the throughputs of all practical ARQ schemes and it serves as a benchmark for gauging the optimality of specific ARQ schemes.
4) We put forth an ARQ design for PNC built upon the standard Selective Repeat ARQ (where out-of-sequence packets are stored at the receiver) with Selective Acknowledgements (SACK). New elements unique to PNC (e.g., *stored-packet tracking*, *non-coupled ARQ*, etc.) are incorporated into the ARQ design. We show that, with proper optimization of ARQ parameters (i.e., window size and ACK frequency), our ARQ design can approach the aforementioned throughput upper bound in 3), even for systems in which ACKs are not error free and ACK overhead is not negligible.

We remark that in this paper, for concreteness and for a focus, the above findings are mainly introduced and studied based on the cross atom. Our findings, however, are applicable to other PNC atoms, as they are related to general principles rather than specificity of the PNC atom structure. Section VII discusses the general applicability of our proposed schemes and our find-ings to other PNC atoms. Furthermore, the same section explains that these design principles can also be applied when "straightforward network coding" (SNC)[3] rather than PNC is used.

---

[2] We give a rigorous analytical proof that non-coupled ARQ is more efficient than coupled ARQ in the cross atom. Since the analytical proof for other atoms is more complex to present, we verify the other atoms by simulations. However, in principle the same proof technique can be used for the other atoms.

The rest of this paper is organized as follows. Section II overviews related work. Section III presents our system model. Section IV analyzes the benefit of storing and tracking coded and overheard packets. A Markov framework is constructed to derive the idealized throughput of a simple ARQ model in which the transmissions of ACKs are error-free and ACK overhead is negligible. Section V shows that non-coupled ARQ is more efficient than coupled ARQ in terms of their idealized throughputs. Section VI investigates the design of Selective Repeat ARQ with SACK in PNC. We show that the idealized throughput of non-coupled ARQ can be approached to within 4% even when ACK transmissions are not error-free and ACK overhead is not ignored. This means that non-coupled ARQ should also have better throughput performance than coupled ARQ in a practical network setting. Section VII discusses the general applicability of our design principles. Section VIII concludes this paper.

## II. RELATED WORK

Although there has been no past work studying the ARQ design for PNC with overheard information, there has been work [4-14] on the ARQ design for SNC with overheard information. However, as elaborated below, certain aspects of PNC raise new issues in the ARQ design that were not addressed in the prior work on SNC ARQ.

Katti et al. proposed COPE [4], a network relaying structure employing wireless network coding and opportunistic listening, i.e., overhearing. COPE uses SNC. Going back to the cross atom, if SNC based on COPE were adopted, nodes A and B would first separately transmit packets *A* and *B* to the relay R in different timeslots. Then relay R would decide to transmit the coded packet $A \oplus B$, the native packet *A*, or the native packet *B*, depending on the reports from nodes C and D on the packets they have previously overheard. In contrast, for PNC, nodes A and B would transmit packets *A* and *B* together in the same timeslot to relay R. Relay R only attempts to decode $A \oplus B$, and not packet *A* or packet *B*. One main concern in SNC with COPE is how the coding node (relay) can be efficiently informed of the packets available at the receivers (i.e., consideration of the reporting overhead). In [5], G.S. Paschos et al. studied a deterministic system, where the contents of the receivers are announced to the coding node via reports, and a stochastic system, where the coding node makes stochastic control decisions based on statistics without explicit reports. An ARQ scheme using NACK was proposed in [6]. In the context of [5], the scheme in [6] is basically a deterministic system. Unlike [4] and [5], where the reports from a receiver are sent to the coding node periodically, in [6] a report (NACK) is triggered by the event that a coded packet has been received and yet the receiver cannot extract its desired packet out of the coded packet.

---

[3] SNC performs network coding at the higher layer. For example, applying SNC to the cross atom, nodes A and B should separately send their respective packet to relay R in two timeslots; in the third timeslot, relay R directly XORs packet *A* with *B*, then broadcasts packet $A \oplus B$.

Sorour et al. introduced the term Instantly Decodable Network Coding (IDNC) [7-9], a network-coded retransmission scheme by constructing IDNC graphs based on feedback from the end users. Lu et al. further proposed to apply IDNC to relay-aided broadcast systems [10] and [11].

The ARQ studies for network coding without exploiting overheard information were proposed in [12-14]. In particular, random linear network coding is employed [12]. The sender transmits multiple random linear combinations of all packets currently in the buffer for broadcast purposes without knowing the stored packets at the receiver. Ref. [14] only studies ARQ for TWRC.

Notably, besides all employing SNC, the prior ARQ studies [4-13] adopt link-by-link ARQ. In particular, these SNC studies formulate the problem as how a base station (relay) finds an optimal encoding scheme based on feedback from the users (destinations) (on both side information and desired packets available to them). The goal of the encoding scheme is to minimize the number of transmissions required to deliver the packets to all users (that is, the index coding problem [15]); or maximize the number of packets that can be decoded at all users after one broadcast [5-11]; or satisfy other requirements. In this sense, the SNC ARQ focuses on the downlink phase in the context of our study here. In particular, it is a one-hop ARQ and it is relay-centric: the relay makes decision, retransmits and collects the feedback. In the context of the cross atom, for example, SNC ARQ investigated the transmissions between the relay R and the destinations C and D. For SNC, the relay can only broadcast a network-coded packet $A \oplus B$ to the destinations only if it has successfully received the native packets $A$ and $B$ from nodes A and B separately. After the relay broadcasts packet $A \oplus B$, if a destination, say node C, fails to extract its desired packet $A$ from packet $A \oplus B$ due to its failure to overhear packet $B$ needed for the extraction during the uplink phase, it sends a request to the relay (which is one-hop away) to directly retransmit the desired native packet $A$. This is because by SNC, the relay already has native packets $A$ and $B$ and it does not make sense for source node A (which is two-hop away) to retransmit packet $A$.

Our PNC ARQ designs in this paper are different because the relay does not have the native packets $A$ and $B$ even if it could decode the XOR packet $A \oplus B$ from the simultaneous transmissions of packets $A$ and $B$ (i.e., the relay is designed to decode the XOR packet). Thus, if destination C in the cross atom, for example, receives packet $A \oplus B$, but fails to overhear packet $B$, and as a result fails to obtain its desired packet, packet $A$, it cannot ask the relay to retransmit packet $A$, since the relay does not have it. Although PNC has an advantage over SNC in that it allows simultaneous transmissions in the uplink, the fact that the relay does not have the native packets even if it can obtain the XOR packet means that it cannot perform link-by-link ARQ as SNC does in the above example. Retransmissions of native packets will have to be initiated by the sources, and not the relay. To simplify design, our paper here focuses on end-to-end ARQ, where the relay is oblivious of the ARQ mechanism, and the ARQ mechanism is between the sources and the destinations.

Compared with link-by-link SNC ARQ, where the uplink ARQ is decoupled from that of downlink ARQ, our PNC ARQ has to consider both uplink and downlink phases as a whole: the source nodes collect feedback, make decision, and retransmit if necessary. This additional constraint in PNC makes the ARQ design somewhat more challenging. In particular, besides satisfying the request from the targeted destination, the packet retransmission from a source node also allows other destination nodes to re-overhear the missing overheard packets. Therefore, for better performance, it behooves a destination in PNC ARQ to store and track the coded packets that cannot be decoded due to missing overheard packet, in case the side information can be re-overheard in the future. By contrast, for the link-by-link SNC ARQ, as far as the source is concerned, as soon as its packet reaches the relay correctly and is ACKed by the relay, it will not retransmit the packet again, thus re-overhearing of this packet is not possible. As a result, there is no need to store and track the coded packets that cannot be decoded in the link-by-link design. In short, the different roles of the relay in PNC and SNC, and the possibility of re-overhearing of the side information make the ARQ for PNC different from the link-by-link ARQ for SNC.

An additional notable point related to the ARQ studies [4-14] is that they assumed error-free ACK (except [8]) and did not consider feedback overhead due to ACK. In practical wireless systems, ACK packets can be lost in transmission and sending ACK packets does consume time and can cause throughput degradation. From our analysis in this paper, such throughput degradation cannot be neglected. We investigate how to minimize throughput degradation by optimizing the transmission window size and ACK frequency. When the ACK is lost, unlike [8] which adopts a probabilistic way to estimate the status of the receiver, we adopt a deterministic way to retransmit the packet that is not ACKed.

Some ARQ implementation issues for PNC systems were proposed in [16] and [17]. However, they only consider PNC in TWRC, where there is no overhearing. As will be seen later in this paper, it is the possibility of overhearing that adds new angles to the PNC ARQ design.

Besides ARQ, another way to realize error control in communication systems with noise is forward error correction (FEC) via channel coding. Overview of issues related to PNC channel coding can be found in the tutorial paper [18]. As with ARQ, there are two possibilities for FEC: end-to-end and link-by-link channel coding. In the end-to-end approach [19], channel coding is transparent to the network-coding system. That is, channel coding can be considered as being applied at an upper layer above PNC system at only the end nodes. In the link-by-link approach [20-29], channel coding and network coding functionalities can be integrated together for better performance. Such link-by-link channel-coded PNC allows the relay to denoise the signals before forwarding the network-coded message along. A number of these papers also consider the asynchrony issues when the signals of multiple simultaneously transmitting nodes arrive at the relay with symbol offset, phase offset, and carrier frequency offset [30-32].

Our current paper focuses on the use of ARQ as part of the overall error control mechanism.

The assumption is that despite PHY-layer channel coding (FEC), packets and acknowledgements can still encounter errors and be lost. ARQ is a means to deal with the remaining post-FEC errors. We also assume that the PHY-layer asynchronies, such as symbol offset, phase offset, and carrier frequency offset, have been dealt with at the PHY layer [30-32] and that the ARQ design at the higher layer does not have to consider them anymore.

### III. SYSTEM MODEL

As shown in [2], a general relay network operated with PNC can be designed by decomposing it into a number of PNC building blocks. We refer to these building blocks as *PNC atoms* because they cannot be further decomposed into smaller building blocks. Broadly speaking, a PNC atom specifies how the traffic among a subset of nodes around a relay can be delivered via the relay using PNC.

In this paper, for concreteness, we will first focus on a specific PNC building block with the "cross structure" as shown in Fig. 1, as has already been introduced in the introduction section. We choose to focus on the cross atom for three reasons: 1) It is an atom that utilizes overheard information. As explained in [2], there are two types of atoms, those that utilize overheard information, and those that do not. The ARQ design of atoms that do not utilize overheard information is similar to the traditional ARQ design. Only atoms that utilize overheard information will face new subtleties as addressed in this paper. 2) It is one of the three atoms that are most important for ensuring good performance [2]. In particular, in optimizing the transmission schedule by the method of network decomposition described in [2], these three atoms are used most often in the optimized solution; omitting the other atoms results in little throughput degradation. The other two important atoms are the TWRC atom, which exploits self-information but not overheard information; and the "Special" TWRC atom [2], which only utilizes overheard information in one flow. Since the cross atom utilizes overheard information in two flows and is more general, we can directly extend the ARQ design for the cross atom to the special TWRC atom. 3) The general principles of the ARQ design derived from the cross atom also apply to all other atoms that utilize overheard information. We discuss the general applicability of the ARQ principles in Section VII.

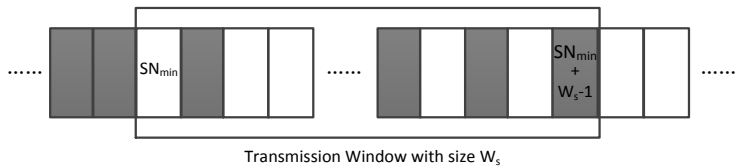

Fig.2 An example of the transmission window.
The grey boxes represent packets that have been acknowledged. The white boxes represent packets that have not been acknowledged.

**Traffic Flows:** We consider multiple unicast traffic traversing the same relay. As shown in the traffic flow diagram of Fig.1, node A has a stream of packets, $A_i$, $i$=1, 2, …, destined for node C; node B has a stream of packets, $B_i$, $i$=1, 2, …, destined for node D. For each of the two flows, the destination (e.g., node C) is outside the transmission range of the source (e.g., node A). The flows rely on relay R to assist the delivery packets to the destinations. As indicated in Fig. 1, the destination of each flow is within the transmission range of the source of the other flow, and can therefore overhear the transmission that source. Specifically, node C can overhear node B, and node D can overhear node A.

**Transmission Window:** To ensure in-order delivery of packets, we set a "window" on each source. The window covers a sequence of $W_s$ packets that can be transmitted by the source. All packets to the left of the left boundary of the window have been successfully delivered to the destination. The source knows this through the ACKs fed back from the destination. The left boundary marks a packet that has not been acknowledged by the destination. The sequence number of the left-boundary packet is $SN_{min}$, as shown in Fig. 2. The right boundary contains the "largest packet" (packet with the largest sequence number $SN_{min} + W_s -1$) that could have been sent by the source at the moment in time. The source cannot send packets to the right of the right boundary without first receiving an ACK indicating that packet $SN_{min}$ has been received at the destination (this acknowledgement will advance the left boundary and the right boundary of the window, hence allowing more packets to the right to be sent). In general, the window may contain (i) some packets that have been sent and acknowledged to have been received correctly (note that packet $SN_{min}$ does not belong to the group of packets whose receptions have been acknowledged), (ii) some packets that have been sent and not yet acknowledged, and (iii) some packets that have not been sent but may be sent shortly by the source. The source sends the packets in the window in the order of their sequence numbers. When it reaches the right boundary of the window, it wraps back to the left bound of the window and retransmits $SN_{min}$. Only unacknowledged packets within a window will be retransmitted.

**Link Success Probability:** In an error-prone network, packet transmissions on a link may fail. We define *link success probability* (*LSP*) as the expected number of correctly received packets divided by the total number of transmitted packets—coded as well as native packets—on a link. In other words, *(1-LSP)* is the packet loss probability on a link. In PNC systems, there are two types of links: direct link (e.g., link R-D in Fig.1) where the receiver is the destination of the transmitted packets; and overhearing link (e.g., link A-D in Fig.1) where the receiver is not the destination of the transmitted packets. Note that the link from which the relay gets the XORed packet from the superimposed signals from the sources is also counted as a direct link (its LSP is denoted as $p_1$ in Fig.1).

**Transmission pattern:** A transmission pattern is a sequence of successive transmissions by nodes in the building block to deliver one packet of each traffic flow to its destination. As shown in Fig.1, the transmission pattern of the cross structure consists of two time slots. In each time slot, some nodes transmit and some nodes receive. The packets transmitted by the transmitting nodes and the packets received by the receiving nodes are specified in the transmission pattern in Fig. 1.

**Round:** In a PNC atom, if the flows want to deliver a stream of packets, the sequence of transmissions in its transmission pattern will be repeatedly scheduled. Each sequence of transmissions in the transmission pattern is called a *round*. Since a transmission pattern always starts with some source nodes transmitting their packets to the relay, and each source node always transmits once in a transmission pattern, each time the source nodes begin to transmit a new packet or retransmit an old packet, we say that a new *round* begins. For example, in Fig.1, each time nodes A and B transmit to relay R, they start a new round. If the network is error-free, after one round, each flow successfully delivers one packet. If the network is not error-free, multiple rounds may be needed for the delivery of one packet. This paper focuses on error-prone networks.

**Coupled ARQ**: In an atom operated with coupled ARQ, its multiple unicast end-to-end flows will not deliver their next packets until all the flows have successfully delivered their current packets. In a round, if any flow fails to deliver its current packet, the transmission pattern will repeat for all current packets of all sources in the next round. For example, with respect to the cross atom, when node A is retransmitting $A_i$ (indexed by $i$), node B must be transmitting $B_i$ (indexed by the same $i$), even if $B_i$ has been received by node D in the previous round. When all packets in the current round are delivered successfully, the multiple flows will then begin to transmit their next packets in a new round. However, there is still an outstanding subtlety. Specifically, it is possible for node A to receive ACKs indicating that both $A_i$ and $B_i$ have been received by their destinations, while node B either did not receive both ACKs or did not receive one of the ACKs. In this case, node A may progress to transmit packet $A_{i+1}$ while node B will still transmit packet $B_i$ in the next round. This causes decoupling of the transmission. In this paper, we make an idealized assumption that somehow nodes A and B know what ACKs the other node has received (e.g., through a backend network or by other means). In the above example, since node B also knows the ACKs of node A, node B will also progress to transmit packet $B_{i+1}$ in the next round. We will show that despite this artificial advantage given to coupled ARQ, it still does not perform as well as non-coupled ARQ.

**Non-coupled ARQ:** For non-coupled ARQ, each unicast flow will deliver its next packet after the reception of its current packet has been acknowledged, regardless of the outcomes of other flows.

**End-to-end ARQ**: This paper considers end-to-end ARQ designs. In end-to-end ARQ, the relay does not participate in the ARQ process. Specifically, the feedback (ACK) is provided by destination nodes only. The relay only forwards the ACK from the destination nodes to the source nodes. In addition, only the source nodes retransmit packets; the relay does not store packets for retransmission purposes.

**ACK frequency and feedback mechanism**: ACK frequency is the rate at which ACK is returned, expressed as the number ($N$) of Data packet *receptions* per ACK. Here, one *reception* means a coded packet that contains a native packet destined for the destination is successfully received from the relay (regardless of whether the desired packet can be extracted from the coded packet). After every *N* receptions, the destination will send back an ACK to the relay.[4] ACKs, if any, are always returned at the end of a round. After receiving the ACKs from the destinations, the relay combines all received ACKs into one ACK and broadcasts it to their sources (hence, all sources can also know the statuses of others) before the next round. Note that, when there is an ACK, it is always sent at the end of a round before the next round. Thus, when an ACK is received at the source, it contains the most updated status at the destination at the end of a round. In other words, the reception of an ACK immediately aligns (synchronizes) the window statuses at the source and the destination (i.e., with the selective repeat and select ACK mechanism investigated in this paper (in Section VI), the knowledge on packets have been successfully received at the source is consistent with that at the destination immediately after an ACK is received).

## IV. STORED-PACKET TRACKING AND THROUGHPUT ANALYSIS

We explained the importance of *store-packet tracking* by means of an example in Section I. Specifically, the destination node should track coded packets and overheard packets, so that it can extract its desired packets from coded packets and overheard packets received from different round. This section formally presents this tracking design and constructs a Markov model to analyze the resulting throughput.

Since the tracking design for non-coupled ARQ is more intricate than that of coupled ARQ, and non-coupled ARQ is more efficient than coupled ARQ (to be shown in Section V), we focus on the stored-packet tracking design for non-coupled ARQ here.

### A. Stored-Packet Tracking

For PNC systems with overheard information, such as the cross atom, successful extraction of a packet at the destination depends on the availability of both the coded packet and an associated overheard packet. It is possible that, in some rounds, the destination node only receives either the coded packet or the overheard packet. Although the packet extraction will fail for that round, the received packet can be stored for future use. (This is the case for both coupled and non-coupled ARQs.) In our scheme, the destination nodes separately store the overheard packets and the coded packets in an *overheard-packet pool* (*O-pool*) and a *coded-packet pool* (*C-pool*), respectively.

Since each native packet can be retransmitted by the source multiple times before it is suc-cessfully extracted at the destination, and different flows are not coupled, a packet $A_i$ that is retransmitted in a round may be network-coded with a packet $B_j$ where the index $j$ may not be the same as $i$. In particular, an overheard packet may be used to extract desired packets from different coded packets. For example, it is possible for node D to receive $A_i \oplus B_i$, $A_i \oplus B_j$ and $A_i \oplus B_k$ in three different rounds. These three coded packets contain the same $A_i$, thus if node D overhears $A_i$ in some round, it can then successfully extract $B_i$, $B_j$ and $B_k$ from the three coded packets. In other words, if in a round, the destination node only receives a coded packet but fails to receive the overheard packet, the coded packet may still be decoded by a side packet overheard in a past or future round. Proper *stored-packet tracking* requires that each time the destination overhears a packet (e.g., $A_i$), it checks all the stored coded packets (e.g., $A_i \oplus B_i$, $A_i \oplus B_j$ and $A_i \oplus B_k$) in the *C-pool* to see whether some desired packets can be decoded using these packets and the overheard information just received. By the same token, each time the destination receives a coded packet (e.g., $A_i \oplus B_k$), if it fails to overhear the corresponding packet in the same round (e.g., $A_i$), it should track the stored overheard packets (e.g., $A_i$) in *O-pool* to see whether an available overheard packet can be used together with the coded packet to extract a desired packet.

---

[4] A number of different schemes for ACK delivery are possible in PNC. One possibility is as follows. In the case of the cross atom, or atoms with only two destinations, if the two destinations transmit ACKs simultaneously, the PNC mechanism can also be applied at the relay to get a network-coded ACK before forwarding the network-coded ACK to the sources. More generally, in a general atom with more than two destinations, if more than two destinations send their ACKs simultaneously, to keep things simple, the relay can treat that as a "collision" (after all, even without collisions, ACK could be lost). In particular, the collision probability decreases with the increase of ACK frequency parameter *N*. For example, for a Star atom with three destinations (see Section VII and [2]), simulations indicate that for *N*=4 and *LSP*=0.8, the ACK collision probability is only 3%.

Note that since an overheard packet (e.g., packet $A_i$ received by node D in the above example) can be used to decode multiple coded packets, even if the extraction of the desired packet is successful in the round when the overheard packet is received (e.g., node D also receives the coded packet $A_i \oplus B_i$ so that node D can extract packet $B_i$), it will still be useful to store the overheard packet (e.g., for node D to store $A_i$) for potential future use. This is because the target destination of the overheard packet may not have received the packet in this round, and that the source of the overheard packet will need to retransmit the overheard packet (e.g., node A retransmits $A_i$ because node C did not receive it in this round). When the source retransmits the packet, it may be network-coded with yet another packet by the relay (e.g., while node A retransmits $A_i$, node B transmits $B_j$ in a future round so that the relay forwards $A_i \oplus B_j$). The stored overheard packet can then be used to extract a desired future packet even if this future retransmission fails to be overheard (e.g., node D cannot receive $A_i$ in this future retransmission, but node D already has $A_i$ stored from a previous round; node D can then use the stored $A_i$ and the received $A_i \oplus B_j$ to extract $B_j$).

Another issue of stored-packet tracking then arises as to when a stored packet can be removed from *O-pool* or *C-pool*. Before a packet is removed, we need to ascertain that it will be no longer useful in the future. Let us first consider the rule for removing packets from the *C-pool*. In general, a coded packet in *C-pool* can be removed if the desired native packet embedded in it has been successfully extracted (e.g., node D would drop $A_x \oplus B_i \; \forall \; x$ where $B_i$ has been extracted). This is because the coded packet will no longer be useful for the extraction of the desired packet given that the desired packet has been obtained. Now, let us consider the *O-pool*. In general, an overheard packet in the *O-pool* can be removed as soon as the packet has been successfully delivered by its original flow (e.g., packet $A_i$ can be dropped from the *O-pool* of node D as soon as the successful reception of packet $A_i$ is acknowledged by node C)[5]. This is because the packet will not be transmitted by its source node anymore and hence no future XOR packet will embed the packet (e.g., no future XOR packet relayed by the relay will be of the form $A_i \oplus B_x \, \forall \, x$).

Overall, ***stored-packet tracking*** is a mechanism that not only keeps track of which stored packets can be used for extraction, but also keeps track of which stored packets can be discarded because they are no longer useful. This is summarized as the first mechanism of our PNC ARQ :

**Mechanism 1**: PNC ARQ adopts ***stored-packet tracking*** to increase the throughput.

To show the advantage of the mechanism, in subsection B, we investigate the extent to which tracking can increase throughput. Then in the subsection C, we demonstrate the throughput gain of tracking relative to the original system.

*B. Throughput Markov model for non-coupled ARQ with window size 1*

In general, the throughput analysis of an ARQ system with tracking can be rather complex. Without packet storage and tracking (i.e., unused packets received in a round are immediately discarded), the throughput of the cross atom is $p_1(p_2p_4+p_3p_5)$ packets per round, where $p_i$ are the LSPs (link success probabilities) on the respective links as indicated in Fig. 1.

With packet tracking, the throughput is not immediately apparent. To study the throughput with stored-packet tracking, here we provide a closed-form analysis for an idealized ARQ model. We construct a Markov model for this analysis. The throughput of the idealized ARQ model serves as an upper bound for the throughputs of more realistic ARQ models. Later in Section VII, we show by simulations that the throughput of the realistic ARQ (in our design) can be very close to the throughput of the idealized ARQ.

The following assumptions are made in the ***idealized ARQ***: (1) immediate error-free ACK feedback, (2) negligible ACK transmission time. With the ideal model, each source immediately knows whether the current packet has been successfully delivered at the end of each round.

To reduce the number of states for analytical tractability, we also assume that the transmission window size of each flow is one, that is, the source node will keep on sending the same packet until this packet has been successfully delivered. Intuitively, under the ideal model, this assumption of window size being one will not affect the decoding probability, hence the throughput (i.e., assuming window size larger than one will not improve the throughput in the ideal model). For example, in the cross atom, suppose that node D overhears packet $A_i$ but $A_i$ is not success-fully delivered to node C. Thus, node A will retransmit $A_i$ in a future round (if the window size is one, then it is in the next round; if the window size is larger than one, the retransmission may not happen immediately in the next round). In a future round, if node D receives a packet $A_i \oplus B_j$, it can extract $B_j$ by exploiting the stored $A_i$. Note that $B_j$ can be any packet in the window. For larger window, there are more possible choices of packet $B_j$ that could be transmitted simultaneously with packet $A_i$. However, which packet $B_j$ is paired with $A_i$ does not affect the probability of receiving packet $A_i \oplus B_j$ at the destination nodes. Thus, in that sense, having a window size larger than one will not improve the probability of delivering a coded packet. Neither will the probabilities of overhearing packets $A_i$ and $B_j$ be improved. For another example, suppose that node D receives $A_i \oplus B_j$ but misses the overheard packet $A_i$. So long as $A_i$ is not successfully delivered to node C (hence $A_i$ should be retransmitted), node D always has the same probability to re-overhear $A_i$ in some future round, hence can decode the stored $A_i \oplus B_j$. The larger window size just means $A_i$ would be retransmitted later but larger window size does not improve the overhearing probability of $A_i$. To validate our intuition, we have performed massive simulations based on different window sizes. The results show that their throughputs are same regardless of window size (we omit the results here to conserve space).

---

[5] As mentioned in Section III, since the relay would broadcast ACKs, each destination can overhear the statuses of other flows.

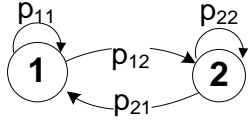

Fig. 3. Macroscopic Markovian States 1 and 2

For simplicity, we also assume in our analysis that the LSP of different direct links are homogeneous (denoted by $p_1$) and the LSP of different overhearing links are also homogeneous (denoted by $p_2$). As far as the general principles for PNC ARQ is concerned, we find that the conclusions obtained from the analysis of the homogeneous study in this paper also apply to the inhomogeneous case generally. Although a rigorous proof is not available due to the analytical complexity of the inhomogeneous case, these findings have been validated by our simulation results.

*1. Two Throughput States:*

Since the network is error-prone, there may be some rounds with no successful delivery to either node C or node D. For rounds with successful deliveries, there are two possibilities: 1) only one of the two flows successfully delivers its packet; 2) both flows successfully deliver their packets. Accordingly, we first construct a "macroscopic" Markov chain with two states only (see Fig. 3): **State 1**, the system just delivered one packet (the last successful round delivered one packet); **State 2**, the system just delivered two packets (the last successful round delivered two packets

The transitions in the macroscopic Markov chain may take varying amounts of time, depending on how many unsuccessful rounds there are before the next successful round. Each transition in Fig. 3 may take several rounds until at least one desired native packet is successfully decoded at one of the destinations.

We refer to the two macro states in Fig. 3 as ***throughput states***. The determination of the probabilities of the throughput states, and the average amounts of time the systems stays in the states before the next transition, give us the system throughout, as explained below. With reference to Fig. 3, let $P_s$ be the equilibrium distribution of state $S$, $S \in \{1,2\}$. Let $T_s$ be the average sojourn time in state $S$ before the next transition, measured in number of rounds. A transition is triggered when either one or two packets have been delivered in a round. Note that upon a transition, the next state can be either the same state or the other state. The system throughput (in number of packets per round) is

$$Th_1 = \frac{P_1 + 2P_2}{P_1 \cdot T_1 + P_2 \cdot T_2} \tag{1}$$

2. *Calculating the equilibrium distributions $P_1$ and $P_2$:*

First, we have $P_1 + P_2 = 1$. Second, with respect to the Markov chain in Fig. 3, we have the balance equation: $P_1 \cdot p_{12} = P_2 \cdot p_{21}$, where $p_{ij}$ denote the transition probability from macro state $i$ to macro state $j$.

Next, we need to derive $p_{12}$ and $p_{21}$. We can model a macroscopic transition in Fig. 3 with a transient "microscopic" Markov process, as described below and shown in Fig. 4. A microscopic Markov process with some initial state is launched when a macroscopic transition occurs. The microscopic process then goes through a sequence of microscopic state transitions until it reaches an absorbing state, which triggers the next macroscopic transition. This kicks off another transient microscopic Markov process. Whereas each macroscopic transition may take several rounds, each microscopic transition takes exactly one round.

We first list the possible microscopic states of the microscopic Markov process (note that with window size of one, at a destination, at most one coded packet and at most one overheard packet will be stored):

State ϕ: Both destinations do not have a coded packet or an overheard packet.

State O: One destination has an overheard packet but no coded packet; the other destination has neither.

State X: One destination has a coded packet but no overheard packet; the other destination has neither.

State XO: One destination has a coded packet but no overheard packet; the other destination has an overheard packet but no coded packet.

State OO: Both destinations have an overheard packet but no coded packet.

State XX: Both destinations have a coded packet but no overheard packet.

Upon a macroscopic transition that lands the system in state 2 in Fig. 3, the system will begin at state ϕ at the microscopic level in Fig. 4. This is because both source nodes will begin transmitting new native packets and that all stored information at the destinations is no longer useful. On the other hand, upon entering state 1 in Fig. 3, the system will begin at state O at the microscopic level in Fig. 4, as explained below. When only one native packet is successfully decoded (e.g., packet $A_1$ is decoded at its destination node C, but packet $B_1$ is not yet decoded at its destination node D), the successful destination (e.g., node C) must have an overheard packet (e.g., packet $B_1$) from the unsuccessful source node (e.g., node B). Since the unsuccessful source node (e.g., node B) will still need to retransmit the same packet (e.g., $B_1$) in the next round, and the successful destination (e.g., node C) already has already overheard that packet (e.g., $B_1$), for the next round, the successful destination only need to receive the new coded packet (e.g., $A_2 \oplus B_1$) and will not need to receive the overheard packet again (e.g., $B_1$). Meanwhile, for the unsuccessful destination (e.g., node D), even if it already has an overheard packet from the current round (e.g., $A_1$), this overheard packet can no longer be used for decoding the next coded packet (e.g., $A_2 \oplus B_1$); similarly, even if the unsuccessful destination (e.g., node D) already has the coded packet from the current round (e.g., $A_1 \oplus B_1$), this stored coded packet is no longer useful in the next round. Hence, at the beginning of the next round, the storage of the other destination has neither a useful coded packet nor a useful overheard packet.

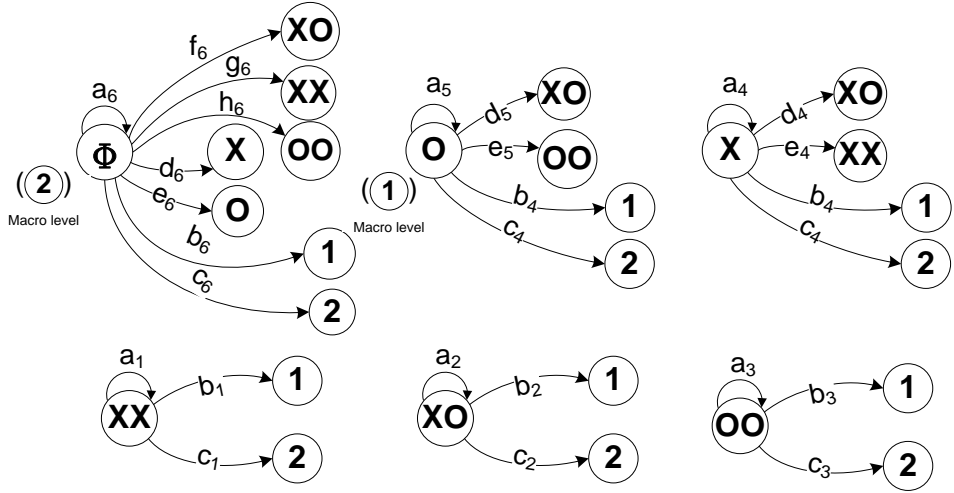

Fig. 4 Transition graphs that show how different microscopic states transit to macroscopic states 1 and 2.
Note that upon entering macroscopic state 1 (2), the system will begin at microscopic state O ($\phi$).

As depicted in Fig. 4, at the microscopic level, the system may visit other microscopic states before the next macroscopic transition is triggered. For example, the system could begin at state O (microscopic), then go to state XO (microscopic), then go to state 2 (macroscopic). In Fig. 4, states 1 and 2 are the absorbing states corresponding to where the system lands at the macroscopic level. We break the graph into several subgraphs to avoid cluttering. It should be understood that these subgraphs should be pieced together for the overall transient Markov chain. For example, state O in the first subgraph is the same state O of the second subgraph.

Each transition in Fig. 4 takes exactly one round. The transition probabilities in the graphs in Fig. 4 can be easily calculated. For example, $p_{X \to XO}$ represents the transition probability from mi-cro state X to micro state XO, that is, the destination with no useful stored packet acquires an overheard packet (with probability $p_2$) but misses the coded packet (with probability $(1-p_1^2)$), and the destination that already has the coded packet fails to overhear the required packet (with probability $(1-p_2)$). So $p_{X \to XO} = p_2 (1-p_1^2)(1-p_2)$.

In Fig. 4, the transient Markov chain will end in one of the absorbing states corresponding to the macro throughput state $S, S \in \{1,2\}$. Let $p_i^{(S)}$ denote the probability that system will end in absorbing state $S$ given that it is now in microscopic state $i$. Since $S \in \{1,2\}$, $p_i^{(1)} + p_i^{(2)} = 1 \ \forall i \in \{1,2\}$ in Fig. 4. Then $p_i^{(S)}$ can be calculated by $p_i^{(S)} = \sum_j p_{i \to j} p_j^{(S)} = \frac{1}{1-p_{i \to i}} \sum_{j \neq i} p_{i \to j} p_j^{(S)}$. Moreover, since upon entering macroscopic states 1 and 2, the system will begin at microscopic states O and ϕ, we can see that $p_{12} = p_O^{(2)}$ and $p_{21} = p_\phi^{(1)}$. Remembering that $P_1 + P_2 = 1$ and $P_1 \cdot p_{12} = P_2 \cdot p_{21}$ (as shown in Fig. 3), we can get $P_1$ and $P_2$ by $P_1 = p_{21}/(p_{12}+p_{21})$ and $P_2 = p_{12}/(p_{12}+p_{21})$ where $P_1$ and $P_2$ are the macro-state stationary probabilities.

3. *Calculating the expected transmission time $T_1$ and $T_2$:*

As mentioned before, not every round leads to a successful delivery (i.e., triggers a new throughput state), after entering a new macro throughput state, it could take several unsuccessful rounds before the transition to the next macro throughput state. To get $T_1$ and $T_2$, we also turn to the microscopic Markov process (in Fig.4) to calculate the expected number of transitions (rounds) starting from an initial state (1 or 2) until an absorbing state (1 or 2).

Let $E[T|i]$ denote the expected number transitions until the system reaches an absorbing state given that the current microscopic state is state $i$. We can write:

$E[T|i] = p_{i \to i} \cdot (E[T|i]+1) + \sum_{j \neq i} p_{i \to j}(E[T|j]+1) = \frac{1}{1-p_{i \to i}} \left( \sum_{j \neq i} (p_{i \to j} \cdot E[T|j]) + 1 \right)$. The detailed equations for different microscopic states $i$ can be derived from the transition graphs in Fig. 4. For example,

$E[T|O] = \frac{1}{1-p_{O \to O}} (p_{O \to XO} \cdot E[T|XO] + p_{O \to OO} \cdot E[T|OO] + 1)$ and $E[T|XO] = \frac{1}{1-p_{XO \to XO}}$. Moreover, since upon entering macroscopic states 1 and 2, the system will begin at microscopic states O and ϕ, we can see that $T_1 = E[T|O]$ and $T_2 = E[T|\phi]$. By solving a set of linear different equations, we can get $T_1$ and $T_2$.

Substituting all the solved variables into equation (1), we get a rather complicated expression of $Th_1$[6], consisting of a complex fraction whose numerator and denominator are both large-degree polynomial functions of $p_1$ and $p_2$, where $p_1$ and $p_2$ denote the LSPs of direct and overhear links respectively.

---

[6] $Th_1=$ $((2p_1^2(p_1^{11}(-1+p_2)^5 - (-2+p_2)^3p_2^4 - p_1^{10}(-1+p_2)^5(6-2p_2+p_2^2) + p_1^9(-1+p_2)^4(-12+17p_2-11p_2^2+4p_2^3) - p_1^8(-1+p2)4(-8-3p2-2p22+2p23)-$
$p_1^7(-1+p_2)^3p_2(-44+76p_2-43p_2^2+10p_2^3) + p_1^6(-1+p_2)^3p_2(-36+57p_2-39p_2^2+11p_2^3) + p_1^5(-1+p2)^2p_2^2(-48+84p_2-45p_2^2+8p_2^3) - p_1^4(-1+p_2)^2p_2^2(-56+102p_2-62p_2^2+13p_2^3) + p_1^3p_2^3(-16+42p_2-39p_2^2+15p_2^3-2p_2^4) + 3p_1^2(-2+p_2)^2p_2^3(3-5p_2+2p_2^2))$
$/$ $(2p_1^{14}(-1+p_2)^3p_2 - p_1^{13}(-1+p_2)^3(1+10p_2+p_2^2) + p_1^{12}(-1+p_2)^3(3+20p_2+3p_2^2) - p_1^{11}(-1+p_2)^3p_2(20+p_2) - p_1^{10}(-1+$
p2313−22p2+17p22−6p23+p24+p19−1+p22−24+64p2−93p22+77p23−31p24+5p25−p18−1+p22−16+29p2−41p22+42p23−2
5p24+5p25−p17−1+p22p266−161p2+147p22−57p23+9p24+p16−1+p22p260−164p2+167p22−79p23+15p24+p15p22−42+14
9p2−205p22+135p23−42p24+5p25+p14p2272−264p2+383p22−275p23+98p24−14p25−p13−1+p22p238−6p2+p22+3p12−2+
p22p233−5p2+2p22−−2+p23p24)

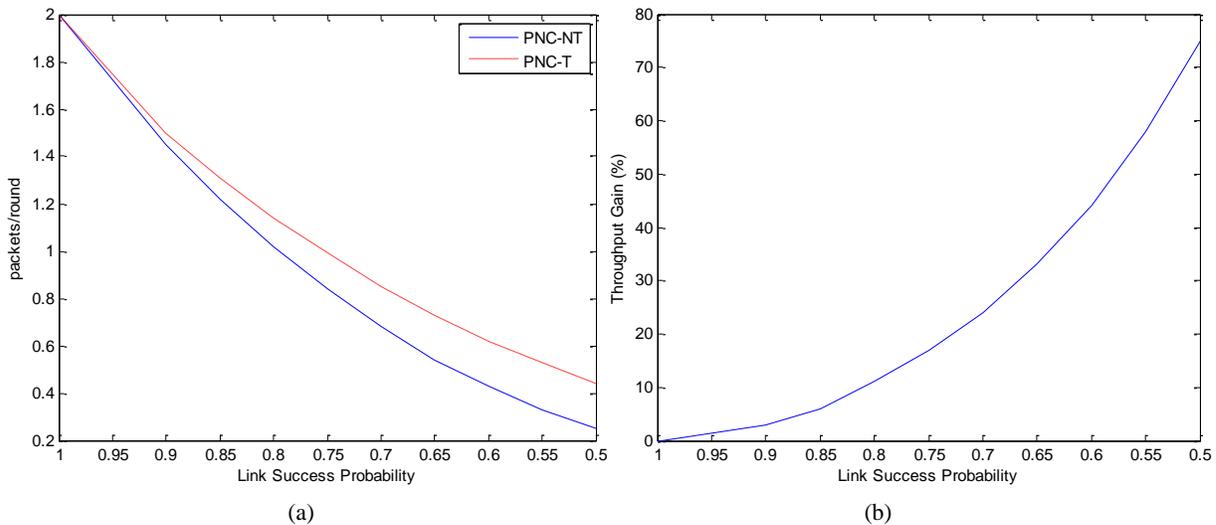

Fig. 5 Throughputs comparison (a) and gain (b) of PNC-T relative to PNC-NT under Different Link success probabilities

## C. Performance Gain of Stored-Packet Tracking

Given the throughput formula $Th_1$ (throughput of non-coupled ARQ with stored-packet tracking) we put forth the following proposition:

**Proposition 1**: adopting *stored-packet tracking* under *non-coupled ARQ* can increase the throughput of the cross atom system with idealized ARQ.

Proof: Without packet storage and tracking, the throughput of the cross atom is $Th_3 = 2p_1^2 p_2$ packets per round (where $p_1$ and $p_2$ denote the LSPs of direct and overhear links respectively). Then, we directly compare $Th_1$ and $Th_3$. Define the throughput difference $f(p_1, p_2)=Th_1 - Th_3$, which is a polynomial function of $p_1$ and $p_2$. Although it is hard to handle the function by hand computation, we can resort to a symbolic computation tool (e.g., Mathematica). We first show that $f(p_{1*}, p_{2*}) > 0$ for some $(p_{1*}, p_{2*})$, where $0 < p_{1*} < 1$, $0 < p_{2*} < 1$ (e.g., $f(0.8, 0.7) = 0.19 > 0$). Then we show the polynomial equation $f(p_1, p_2) = 0$ has no solution[7] for all $(p_1, p_2)$ in the range $0 < p_1 < 1$, $0 < p_2 < 1$. Since $f(p_1, p_2)$ is a continuous function of $p_1$ and $p_2$ (as polynomial function is continuous), the above means that $f(p_1, p_2) > 0$ (i.e., $Th_1 > Th_3$) for all $0 < p_1 < 1$ and $0 < p_2 < 1$. That is, by adopting *stored-packet tracking* in *non-coupled ARQ*, higher throughput can be obtained than without packet storage and tracking. Q.E.D

In Fig. 5, we plot the analytical results of the idealized model to which Proposition 1 applies, where for analytical tractability, ACK loss and overhead have been ignored. Specifically, Fig. 5 shows the throughputs of PNC with tracking (PNC-T) relative to conventional PNC without tracking (PNC-NT) under different LSPs (assuming homogenous link success probability for all direct and overheard links) based on $Th_1$ and $Th_3$. The throughput gain is larger when the wireless channel becomes worse (i.e., in situations where ARQ is needed to guarantee reliability). As shown in Fig.5 (b), the throughput gain increases from 3% to 75% when the LSP decreases from 0.9 to 0.5. A question arises as to whether the throughput gains as seen in Fig. 5 will remain in more realistic networks in which ACK overhead is taken into account and ACK loss can happen. We will examine this issue in Section VI and show by simulations (since the exact analytical studies of the realistic networks are difficult) that indeed the throughput gains of adopting stored-packet tracking in the realistic networks (see Table 4 in Section VI) almost match the direct theoretical calculations of $Th_1$ and $Th_3$ above. Note that we do not list the results when $p$ is too small, because when $p$ is too small, we would be better off using the traditional hop-by-hop transmission scheme. According to the benchmark throughput formulation of PNC, to ensure that PNC is more efficient than traditional hop-by-hop transmission (whose throughput is $p/2$), $p$ must be larger than 0.57.

---

[7] Mathematica can judge whether the polynomial equation has no solution by the Cylindrical Algebraic Decomposition (CAD). Please refer to: http://mathworld.wolfram.com/CylindricalAlgebraicDecomposition.html

## D. More Discussion on Stored-Packet Tracking

The tracking described so far is a single-iteration tracking mechanism. Each time the destination receives an overheard packet (or a coded packet), it checks the stored coded packets in the C-pool (or overheard packets in the O-pool) to see whether a desired native packet can be extracted by combining received packet with a stored packet. However, under window size larger than one, tracking can go further, using multi-iteration tracking. For *multi-iteration tracking*, after an iteration in which a native packet is extracted, it proceeds to another iteration to check if the newly extracted native packet can be combined with a stored packet in the C-pool to extract another packet. This process repeats itself until no more packet can be extracted[8]. For example, suppose that node D has $A_1 \oplus B_1$, $A_1 \oplus B_2$, and $A_2 \oplus B_2$ in its C-pool. When node D overhears $A_2$, it can extract $B_2$ by $A_2 \oplus (A_2 \oplus B_2)$. In the next iteration, it combines $B_2$ with $A_1 \oplus B_2$ to extract $A_1$. Although $A_1$ is not a desired packet, in the iteration after that, node D combines $A_1$ with $A_1 \oplus B_1$ to decode another desired packet $B_1$. Such multi-iteration tracking/decoding can improve throughput.

However, from above example, we can see that only when certain particular combinations of coded packets are stored can multi-iteration tracking/decoding be useful. Our simulation experiments indicate that the additional native packets obtained from multi-iteration tracking (tracking beyond the first iteration) constitute only a small portion of the overall native packets extracted

---

[8] This process is similar to the belief-propagation decoding of LDPC erasure codes.

(less than 3% for different window size, inhomogeneous link success probability and all PNC atoms; we present simulation results under idealized ARQ in Table 1). Thus, there is little need to include *multi-iteration tracking* in our design. In this paper, the single term *tracking* means *single-iteration tracking*.

TABLE 1   ADDITIONAL PACKETS OBTAINED FROM MULTI-ITERATION OVER OVERALL PACKETS UNDER IDEALIZED ARQ

|  | *P=0.9* | *P=0.85* | *P=0.8* | *P=0.75* | *P=0.7* | *P=0.65* | *P=0.6* | *P=0.55* | *P=0.5* |
| --- | --- | --- | --- | --- | --- | --- | --- | --- | --- |
| ***Portion*** | 1.1% | 1.2% | 1.3% | 1.4% | 1.6% | 1.8% | 2.0 % | 2.3% | 2.5% |

## V. COUPLED ARQ VERSUS NON-COUPLED ARQ

We begin this section by stating the second mechanism for PNC ARQ, the use of non-coupled ARQ. We then justify this by analyzing and comparing the throughputs of non-coupled ARQ with coupled ARQ, showing that the former is superior to the latter.

**Mechanism 2**: PNC ARQ adopts ***non-coupled ARQ*** to increase the throughput.

We have derived an idealized throughput of non-coupled ARQ in the last section through a Markov analysis. To show the advantage of the mechanism, in the subsection A, we first calculate the idealized throughput for coupled ARQ using a similar model that adopts the same assumptions as in the last section (window size 1, homogeneous LSP, etc.); then in the subsection B, we show that the idealized throughput of non-coupled ARQ is always larger than that of coupled ARQ.

### A. Throughput of Coupled ARQ with window size 1

In coupled ARQ, even if a source (e.g., node A in the cross atom) already knows that the currently transmitted native packet (e.g., packet $A_1$) has been successfully extracted at its destination (i.e., node C), it may still retransmit this packet (e.g., packet $A_1$) until the other source (e.g., node B) also successfully delivers its current native packet (e.g., packet $B_1$) to its destination (i.e., node D). After all sources have delivered their current packets, they move on to transmit their next packets (e.g., packets $A_2$ and $B_2$). That is to say, for coupled ARQ, with respect to the discussion on non-coupled ARQ in Section IV.B, there is only one macro throughput state: both flows successfully deliver their packets. To calculate the throughput of coupled ARQ, for the cross atom, we can just directly compute the expected number of rounds (denoted by $E[T_N]$) from the initial state $\phi$ when both sources begin to transmit two new packets to the end state 2 when these two packets are successfully delivered to their destinations. There are three immediate states between the initial and end states:

State $1\phi$: One destination already has the desired native packet; the other destination has nothing (neither the coded packet nor the overheard packet).

State 1O: One destination already has the desired native packet; the other destination has an overheard packet but no coded packet.

State 1X: One destination already has the desired native packet; the other destination has a coded packet but no overheard packet.

Then, we can get E[$T_N$] by $\mathrm{E}[T_N] = p_{\phi \to \phi} \cdot (\mathrm{E}[T_N] + 1) + \sum_{i \neq \phi} p_{\phi \to i} (\mathrm{E}[T_N | i] + 1) = \frac{1}{1 - p_{\phi \to \phi}} \left( \sum_{j \neq i} (p_{\phi \to i} \cdot \mathrm{E}[T_N | i]) + 1 \right)$,

where $\mathrm{E}[T_N | i]$ denotes the expected number of rounds needed to go from state $i$ to end state 2.

Finally, we can get $Th_2 = 2/\mathrm{E}[T_N]$. (The complete throughput formula is as complicated as $Th_1$ for the non-coupled ARQ in Section IV.B. We skip it here to save space.)

### B. Comparison of Non-Coupled ARQ and Coupled ARQ

Given the throughput formula $Th_2$, we put forth the following proposition:

**Proposition 2**: *non-coupled ARQ* is more efficient than *coupled ARQ* in the cross atom system with idealized ARQ.

Proof: Let us now compare the idealized throughputs of non-coupled ARQ ($Th_1$) and coupled ARQ ($Th_2$). Define the throughput difference $g(p_1, p_2) = Th_1 - Th_2$, which is a polynomial function of $p_1$ and $p_2$. Then we use the same line of argument as proposed in the proof of Proposition 1 (see Section IV.C) to show that $g(p_1, p_2) > 0$ (i.e., $Th_1 > Th_2$) for all $0 < p_1 < 1$ and $0 < p_2 < 1$. So we can say that non-coupled ARQ is more efficient than coupled ARQ.  Q.E.D

Proposition 2 applies to an idealized network in which ACK loss and overhead are not taken into account, and from $Th_1$ and $Th_2$, we can get that the throughput gains of non-coupled ARQ relative to coupled-ARQ range from 8% to 25% when the LSP decreases from 0.95 to 0.57. A question arises as to whether non-coupled ARQ remains more efficient than coupled ARQ in more realistic networks in which ACK overhead is taken into account and ACK loss can happen. We will examine this issue in the next section and show by simulations (since the exact analytical studies of the realistic networks are difficult) that the idealized throughput of non-coupled ARQ can be approached to within 4% even when ACK transmissions are not error-free and ACK overhead is not ignored (see Table 2 in Section VI). This means that non-coupled ARQ should also have better throughput performance than coupled ARQ in a practical network setting. Thus, we focus on non-coupled ARQ in the next section where we extend our discussion to the non-idealized situation.

## VI. ERROR-PRONE ACK

For the idealized throughput derived in Section IV, we neglected the air time used for sending ACKs (acknowledgement) and assumed that ACKs are error-free. However, in practical wireless systems, sending ACK packets does consume time and ACK packets can be lost in transmission, leading to throughput degradation. In this section, we investigate how to design the ARQ mechanism to allow the throughput of a real system to approach the ideal benchmark throughput $Th_1$ despite ACK errors and transmission overhead.

*A. PNC ARQ*

ACK packets are numbered according to the packets that the receiver has received. ACKs allow the transmitter to adjust its transmission window according to the packets already received by the receiver. If the transmitter receives ACKs that indicate certain packets have been received, the transmitter will not transmit these packets again; the unACKed packets, however, will be transmitted/retransmitted in the future.

When ACKs are lost, the transmitter may retransmit packets already received at the receiver (i.e., there may be a discrepancy between the reception status as perceived by the transmitter and the actual reception status at the receiver). To eliminate such *wasteful packet retransmissions*, we design an ARQ based on Selective Repeat (SR) (where out-of-sequence packets are stored at the receiver) with Selective Acknowledgements (SACK) [33]. Hereafter, we refer to our ARQ design as **PNC ARQ**. PNC ARQ incorporates *stored-packet tracking*, *non-coupled ARQ* and *SR with SACK*.

With selective repeat, the transmitter sends packets within a window in the order of their sequence numbers, skipping those packets whose receptions have been acknowledged by the receiver. When the transmitter reaches the right boundary of the window, it wraps back to the left bound of the window to retransmit packets (referred to as *boundary wrap back*) in the order of the sequence numbers again. Updates as to which packets have been received are performed whenever an ACK is received from the receiver.

With selective repeat, the receiver accepts out-of-order packets. With selective acknowledgement, the receiver sends SACKs specifying which packets within a window have been received and which have not been received. In particular, besides the cumulative ACK indicating that all packets with a sequence number smaller than the ACK number have been received, the selective ACK also indicates the noncontiguous received packets with a sequence number larger than the cumulative ACK number. This allows the transmitter to selectively retransmit packets more efficiently, reducing the likelihood of retransmitting packets already received.

Let us introduce the format of SACK frame in PNC ARQ with an example: {7, 0100101}. The first number indicates that the destination node has received all packets before 7, the subsequent bit map indicates which packets after packet 7 that have been received— the length of the bit map is just *W* (window size) minus 1. In the example, the destination node has received packets 9, 12 and 14. The SACK frame tells the source that packets 7, 8, 10, 11, 13 are yet to be received.

With SACK, when window size is large, although some ACKs could be lost, so long as a subsequent ACK (triggered by the reception of a subsequent packet) is received, the transmitter can update the information of all previously received packets. That is to say, after successfully delivering a packet, as long as the transmitter receives an ACK any time before the transmission process reaches the right window boundary and wraps back to the left window boundary, waste-ful retransmissions can be avoided.

Last but not least, we adopt an ACK-oriented wrapping back mechanism in PNC-ARQ. Specifically, after receiving an ACK and updating the packet information (window), the transmitter will immediately wrap back and retransmit the unACKed packets from the beginning of the window (referred to as *ACK wrap back*), because we want to advance the left boundary of the window as quickly as possible. *ACK wrap back* is different from *boundary wrap back* in that ACK wrap back does not entail wasteful retransmissions of packets that have been received. We define the transmissions in between two consecutive *ACK wrap back* as a *transmission cycle*. If there is a *boundary wrap back* within a *transmission cycle*, there will be wasteful retransmissions. For example, assume window size of five and that the transmitter starts transmitting packets in sequence from packet 1 to packet 5. Suppose that after the transmitter transmits packet 4 and before it transmits packet 5, it receives an ACK {3, 1000}, it will immediately wraps back to retransmit packet 3. This is an ACK wrap back. Note that the ACK {3, 1000} indicates that packet 3 has not been received and packet 4 has been received. Furthermore, the ACK {3, 1000} contains full information on the reception status at the receiver at the moment (packet 4 is received, no retransmission of it is needed). Thus, it makes sense for the transmitter to immediately wrap back to retransmit to make up for the missing packets up to packet 4. In this example, after the ACK wrap back, the right window boundary is packet 7. Say, after this wrap back, if the transmitter does not receive any ACK even after it transmits packet 7, it will wrap back to retransmit packet 3. That is a boundary wrap back. Since packet 3 may have been successfully received (but not successfully ACKed), the retransmission of packet 3 can potentially be wasteful.

Given the ACK mechanism in PNC ARQ, we put forth the third mechanism for PNC ARQ as follows:

**Mechanism 3**: In PNC ARQ, the transmission window size $W$ and the ACK frequency $N$ should be jointly optimized to achieve the idealized throughput.

To show the advantage of the mechanism, in the subsection B, we first clarify why $W$ and $N$ should be suitably chosen; then in the subsection C, we analyze how to find a proper set of $W$ and $N$, and demonstrate through simulations that with such properly chosen $W$ and $N$, the idealized throughput can be achieved with little degradation.

*B. Parameters in PNC ARQ*

In PNC ARQ, two parameters can affect system throughput. The first parameter is the transmission window size $W$. As mentioned in Part A, if the transmission process reaches the right boundary of the transmission window without receiving an ACK since it last started from the left window boundary, the transmitter has to retransmit starting from the left boundary of the window.

In that case, some unACKed packets in the window may have been successfully received, and the retransmissions of them will be wasteful. A larger window size gives a higher probability of receiving an ACK before such boundary wrap back (i.e., less wasteful retransmissions). However, a larger $W$ also means that an ACK will contain more bits (i.e, $W$-1) to indicate the delivery state of each packet in the window. As a result, the system consumes more air time to send back each ACK, leading to a larger feedback overhead (i.e., the mean air time consumed by ACKs relative to the mean air time consumed by Data packets). Since the window size should be bounded[9] in a practical system, we need to find a proper $W$ to trade off between these two effects.

The second parameter is the ACK frequency $N$ (see Section III). To reduce the feedback overhead caused by ACK in our design, the destination node can send an ACK after every $N$ successful receptions from the relay. Larger $N$ leads to smaller feedback overhead. However, given a window size $W$, increasing $N$ would increase the chance of wasteful retransmissions. Hence, the values for $N$ and $W$ must be suitably chosen.

*C. Optimizing Parameters in PNC ARQ*

Assuming homogeneous LSP (link success probability) $p$ for both direct and overhearing links as in Section IV.B, we aim to find a proper set of $W$ and $N$ applicable for all $p$. Let us require that the performance degradation from the benchmark throughput to be less than $e_0$ (e.g., 5%) for any $p>0.57$.[10] This total throughput degradation $e_0$ comprises the feedback overhead degradation $e_1$, where $e_1<e_0$ (e.g., $e_1$=2.5%) and the wasteful retransmissions degradation $e_2$, where $e_2<e_0$ (e.g., $e_2$=2.5%), that is, $e_0= e_1+ e_2$.

**Proposition 3**: Decreasing the window size $W$ while increasing the ACK frequency $N$ reduces the degradation due to ACK overhead in cross atom system with our PNC ARQ.

Proof: We derive the expression of $e_1$, the degradation due to feedback overhead $H$, in terms of $W$ and $N$. For each round, the probability for the destination to receive a coded packet (from the source through the relay) is $p^2$. On average, for each batch of $N$ successful receptions, the source node needs to transmit $N/p^2$ times. Thus, the average overhead due to ACK is $H=\dfrac{T_a}{(N/p^2)T_d}$, where $T_a$ is the transmission time of one ACK packet, and $T_d$ is the transmission time of one Data packet.

An ACK packet contains two parts: one part is the $W/8$ bytes used as the indicators in SACK; the other part is the packet header consisting of $K$ bytes. Suppose that a data packet has $D$ bytes, then we get $e_1$=$H$=$\dfrac{K+W/8}{(N/p^2)\cdot D} = \dfrac{K}{(N/p^2)\cdot D}+\dfrac{W/8}{(N/p^2)\cdot D}$ . (2)

We can see that $e_1$ is an increasing function with $W$ but a decreasing function with $N$.   Q.E.D

---

[9] On the other hand, a larger $W$ also means that the destination will need to buffer more out-of-order packets (since selective repeat ARQ is adopted). Hence, bounding $W$ also helps to limit the memory requirement at the destinations.

[10] Recall from the discussion in Section IV.C that when $p$ is smaller than 0.57, we would be better off using the traditional hop-by-hop transmission scheme. Hence, we only consider $p$ in the interval (0.57, 1).

In general, $e_2$, the degradation caused by wasteful retransmissions relative to total transmissions, also depends on $W$ and $N$. Unfortunately, a closed-form expression for $e_2$ as a function of $W$ and $N$ is difficult to derive and we have not succeeded in doing so. In particular, at the moment of an *ACK wrap back* (see definition in the last paragraph of Part A) the packet in each window slot (except the left boundary slot, which is always an unAcked packet) is either a packet that has been received and acknowledged (referred to as an *RA packet*) or a packet that has not been received (referred to as an *NR packet*). The probability of *boundary wrap back* (see third paragraph of Part A) in the next transmission cycle, which causes wasteful retransmissions, depends on $N$ and the number of *NR packets* (there can be anywhere between 1 and $W$ NR packets). After a *boundary wrap back*, besides *RA* and *NR* packets, another type of packets is possible: the packets that have been received but not yet acknowledged (referred to as an *RNA packet*). Since the source node cannot tell which packets are *NR* and which packets are *RNA*, it would retransmit both types of packets The transmissions of *RNA packets* are wasteful retransmissions. To count the wasteful retransmissions, we need to analyze the probabilities of $W^3$ states (there are $W$ slots in the window and each has three possible states, *RA, NR, RNA*). Although a Markov model can be drawn up for a specific pair of $W$ and $N$, the number of states in the Markov model will vary according to the $W$ and $N$ values. Thus, for each ($W$, $N$), there is a Markov chain, the transitions of which are rather complicated. Overall, the Markov model analysis is intractable.

Lacking a closed-form expression for $e_2$, here we provide a heuristic to find a proper set of $W$ and $N$ to approach the minimal throughput degradation. As $e_1$ can be found in closed form, we first limit the scope of $W$ and $N$ based on $e_1$. We need to determine the appropriate $W$ and $N$ that can make $H$ small. Note that (2) is an increasing function of $p$. To cover all possible $p$, we derive $W$ and $N$ based on the largest possible $p$ ($p=1$). Setting $p=1$ in (2) gives

$$H = \frac{K}{N \cdot D} + \frac{W/8}{N \cdot D} = \frac{1}{ND}\left(K + \frac{W}{8}\right) \qquad (3).$$

Then we have 
$$e_1 = \frac{1}{ND}\left(K + \frac{W}{8}\right) \qquad (4).$$

The above shows the tradeoff between $W$ and $N$ given a target $e_1$. For our heuristic, let $e_1 = e_1^{(1)} + e_1^{(2)}$, where $e_1^{(1)} = K/ND$ and $e_1^{(2)} = W/8ND$ — i.e., we can divide the overall overhead budget into two parts, $e_1^{(1)}$ and $e_1^{(2)}$. Since $e_1^{(1)}$ depends only on $N$ (but not $W$), we can find the required $N$ to meet the overhead target $e_1^{(1)}$. For example, we could set $e_1^{(1)} = 1\%$ and find the $N$ accordingly.

Once $N$ is fixed, $e_1^{(2)}$ then depends only on $W$. In particular, the smaller the $W$, the smaller the $e_1^{(2)}$. However, smaller $W$ would lead to larger $e_2$, the throughput degradation due to wasteful retransmissions. On the other hand, to keep the target $e_1$, from (4), $W$ should be upper bounded by $8e_1ND - 8K$.

Next, let us estimate intuitively how $e_2$ may depend on $W$ (as mentioned earlier, the derivation

TABLE 2  THROUGHPUTS COMPARISON OF BENCHMARK AND PNC ARQ WITH OPTIMAL SETTING

|  | *P=0.95* | *P=0.9* | *P=0.85* | *P=0.8* | *P=0.75* | *P=0.7* | *P=0.65* | *P=0.6* | *P=0.57* |
|---|---|---|---|---|---|---|---|---|---|
| **Benchmark** (*p/r*) | 1.73 | 1.50 | 1.31 | 1.14 | 0.99 | 0.85 | 0.73 | 0.62 | 0.57 |
| **PNC-Opt** (*p/r*) | 1.67 | 1.45 | 1.27 | 1.11 | 0.97 | 0.84 | 0.72 | 0.60 | 0.56 |
| **Degradation** | 3.5% | 3.3% | 3.1% | 2.6% | 2.0% | 1.2% | 1.4% | 3.2% | 2.1% |
| **Overhead** | 3.6% | 3.4% | 3.2% | 2.7% | 2.1% | 1.2% | 1.4% | 3.3% | 3.6% |

TABLE 3  THROUGHPUTS OF PNC ARQ UNDER DIFFERENT SETS OF W AND N

|  | *P=0.95* | *P=0.9* | *P=0.85* | *P=0.8* | *P=0.75* | *P=0.7* | *P=0.65* | *P=0.6* | *P=0.57* |
|---|---|---|---|---|---|---|---|---|---|
| **W=1 N=1** (*p/r*) | 1.43 | 1.14 | 0.91 | 0.72 | 0.56 | 0.43 | 0.33 | 0.24 | 0.20 |
| **W=170 N=4** (*p/r*) | 1.67 | 1.45 | 1.27 | 1.11 | 0.97 | 0.84 | 0.72 | 0.60 | 0.54 |
| **Gain** | 17% | 27% | 40% | 54% | 73% | 95% | 118% | 150% | 167% |

TABLE 4  THROUGHPUTS UNDER DIFFERENT LINK SUCCESS PROBABILITIES

|  | *P=0.95* | *P=0.9* | *P=0.85* | *P=0.8* | *P=0.75* | *P=0.7* | *P=0.65* | *P=0.6* | *P=0.57* |
|---|---|---|---|---|---|---|---|---|---|
| **PNC-NT** (*p/r*) | 1.64 | 1.40 | 1.19 | 0.99 | 0.82 | 0.67 | 0.54 | 0.42 | 0.36 |
| **PNC-Opt** (*p/r*) | 1.67 | 1.45 | 1.27 | 1.11 | 0.97 | 0.84 | 0.72 | 0.60 | 0.558 |
| **Gain** | 1.8% | 3.6% | 6.7% | 12.3% | 17.9% | 25.2% | 33.4% | 42.9% | 56.0% |

*The throughput unit in Tables 3-5 is **p/r**: packets per round.*

of the exact closed-form expression is difficult). On average, the destination sends back an ACK after every $N/p^2$ transmissions. And the probability for the source to successfully receive an ACK is also $p^2$. Here we assume transmitting ACK packet in the reverse link has the same LSP. For example, both LSPs of links R-A and A-R are $p$. On average $1/p^2$ ACKs are needed for the source node to successfully receive an ACK. So on average totally $N/p^4$ data packet transmissions are needed before an ACK is received. This means $W$ should be inversely proportional to $p$. The smaller the probability $p$ is, the larger the window size $W$ needs to be. To cover all possible $p$, we should derive $W$ based on the smallest meaningful $p$ ($p=0.57$, see first paragraph of Part C).

In the following, given fixed $N = K/e_1^{(1)}D$ and $p=0.57$, we derive the proper $W$ through simulation. We first simulated a setting in which $W$ is equal to the upper bound $8e_1ND - 8K$ and get a throughput. After that, we gradually decrease $W$ until the throughput begins to decrease. From massive simulations, we found that when $W$ is not significantly smaller than $8e_1ND - 8K$, the throughput degradation due to wasteful retransmissions is little and not sensitive to the change of $W$. For this region, the throughput depends more on the overhead $H$ and is an increasing function of $W$. (i.e., second item in (2)) decreases. As $W$ is decreased further, there comes a point where wasteful retransmissions become significant and that the maximum throughput is achieved at a $W$ that strikes the right balance between overhead $H$ and wastefulness due to retransmissions.[11]

Our simulation results show that our heuristic is an effective method to find a good set of $W$ and $N$. For example, when $K=30$ bytes (For the ACK frame in IEEE 802.11, $K$ is around 30) and $D=600$ bytes (A conservative data packet size in wireless communication), $N=4$ $W=170$, the worst throughput degradation compared with the upper bound benchmark (calculated from throughput formula $Th_1$ in Section IV) is less than 4% for different $p$. Further details on throughput comparison with the benchmark can be found in Table 2, where PNC ARQ with optimal set of $W$ and $N$ is referred to as PNC-Opt. Hence, with proper $W$ and $N$, we almost compensate for the loss due to the ACK lost and the ACK overhead. We also present the total overhead of our PNC ARQ in Table 2. We calculate the overhead by $(2/Th_a - 2/Th_1)/(2/Th_1)$, where $Th_a$ is the throughput of PNC-Opt, $Th_1$ is the benchmark throughput. The overhead presents the extra time slot (in percentage) needed to deliver one packet under the realized network with our PNC ARQ.

---

[11] Note that to find the proper set of $W$ and $N$, we can also first fix $W$ and find the lower bound of $N$ from (4); then directly increase $N$ to reduce the overhead $H$ until the proportion of wasteful retransmissions is large enough. However, intuitively, $N$ is always much smaller than $W$, and the step size of changing $N$ may be too coarse to approach the optimal throughput.

On the other hand, for the simplistic system in which the transmission window $W$ is 1, and in which the receiver sends back an ACK upon each received data frame, the throughput degradation is very significant. In Table 3, we show the throughput gains of *PNC ARQ* with $W$=170 and $N$=4 (i.e., PNC-Opt) relative to *PNC ARQ* with $W$=1 and $N$=1 under different $p$. The throughput gains increase with the decrease of $p$. When $p$=0.57, the gain is 167%.

Table 4 shows the throughput gains of PNC-Opt relative to PNC-NT. Recall that PNC-NT is the conventional PNC without stored-packet tracking. Here, for PNC-NT, we assume error-free ACK and neglect ACK air time. With this assumption, we are giving PNC-NT an advantage over PNC-O because we do not make the same assumption for PNC-Opt. We note that the throughput gains are nearly the same as those in Table 1 (Section IV), which compares PNC-T (with stored-packet tracking—as in PNC-Opt—but also assuming error-free ACK and neglecting ACK air time) with PNC-NT. In conclusion, with proper $W$ and $N$, the throughput gain from stored-packet tracking can be maintained in practical wireless systems with error-prone ACK and with ACK overhead.

## VII. GENERAL APPLICABILITY OF ARQ MECHANISMS

Recall from Section III that, for concreteness, this paper focuses on the cross atom in the ARQ design and analysis. The principles behind the ARQ mechanisms, however, are not limited to the cross atom only. In particular, they apply to all other atoms that exploit overhearing.

In this section, we discuss the general applicability of the ARQ mechanisms proposed in Sections IV to VI for other atoms. Our PNC ARQ contains three key ingredients in its design: 1) *stored-packet tracking*, 2) *non-coupled ARQ*, and 3) optimization of *W* and *N*. We will discuss whether these three design ingredients can be applied to PNC atoms with overheard information (OPNC), PNC atoms without overheard information (Non-OPNC), SNC[12] atoms with overheard information (OSNC), and SNC atoms without overheard information (Non-OSNC). When we say that a mechanism applies to a system, we mean the mechanism can increase the throughput of the system. Table 5 summarizes the general applicability of the three mechanisms for various set-ups.

---

[12] For each PNC atom, there is a corresponding SNC atom. They have the same structure, but different transmission patterns: one utilizes PNC, one utilizes SNC. Generally, an SNC atom consumes more timeslots than its corresponding PNC atom.

TABLE 5    GENERAL APPLICABILITY OF ARQ MECHANISMS

|  | OPNC | Non-OPNC | OSNC | Non-OSNC |
|---|---|---|---|---|
| **Stored-Packet Tracking** | Yes | N/A | Yes | N/A |
| **Non-coupled** | Yes | N/A | Yes | N/A |
| **W and N Optimization** | Yes | Yes | Yes | Yes |

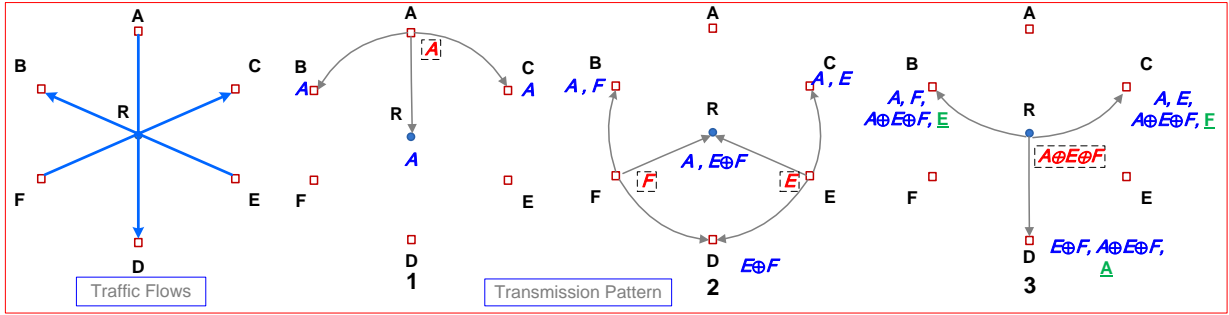

Fig. 6 PNC Star Atom

In the following, to illustrate the general applicability of the ARQ mechanisms to PNC atoms other than the cross atom of focus so far, we first consider the PNC atom as shown in Fig. 6, referred to as the star atom. Fig.6 shows how the star atom consumes three timeslots to deliver one packet for each of three flows. Since the figure is self-explanatory, we will not go into the details here. The ARQ mechanism for the star atom will be explained in Part A below.

*A. General Applicability of Stored-Packet Tracking*

Stored-packet tracking is a design that utilizes overheard information and is not applicable to atoms without overheard information.

*A.1 General Applicability of Stored-Packet Tracking to OPNC*

Let us look at how stored-packet tracking can be applied to the star atom. With respect to Fig. 6, without loss of generality, we focus on destination D. Consider a particular round. Suppose that destination D only overhears $E_i \oplus F_i$ in the second time but fails to receive $A_i \oplus E_i \oplus F_i$ in the third time slot in the 3-slot transmission pattern shown in Fig. 6, proper tracking (see Section IV.A) requires destination D to check whether $A_i \oplus E_i \oplus F_i$ has been stored in its coded-packet pool (note: $A_i \oplus E_i \oplus F_i$ could have been received in a previous round). By the same token, if destination D only receives $A_i \oplus E_i \oplus F_i$ but fails to overhear $E_i \oplus F_i$ in a round, for proper tracking it should check whether $E_i \oplus F_i$ has been stored in its overheard-packet pool (see Section IV.A). As long as the unsuccessful packet (i.e., $A_i \oplus E_i \oplus F_i$ or $E_i \oplus F_i$ in the above) can be tracked and traced back to a previous successful reception of that packet, the desired packet $A_i$ can be extracted. In general, all successfully received packets should be stored for potential use in the future. With our proposed stored-packet tracking mechanism (see Section IV.A), a systematic method is used to determine how long the packets should be stored so that they can be discarded when they are no more useful for extracting desired packets in the future.

TABLE 6    SIMULATION RESULTS UNDER COUPLED ARQ

| LSP | | P=0.9 | | | P=0.85 | | | P=0.8 | | |
|---|---|---|---|---|---|---|---|---|---|---|
| Scheme | | PNC-NT | PNC-T | Gain | PNC-NT | PNC-T | Gain | PNC-NT | PNC-T | Gain |
| A T O M | II | 0.67 | 0.70 | 5% | 0.56 | 0.60 | 7% | 0.46 | 0.51 | 11% |
| | V | 0.62 | 0.66 | 6% | 0.50 | 0.56 | 12% | 0.40 | 0.48 | 19% |
| | VI | 0.65 | 0.69 | 6% | 0.48 | 0.54 | 12% | 0.36 | 0.43 | 20% |
| | VII | 0.35 | 0.38 | 8% | 0.25 | 0.30 | 20% | 0.18 | 0.24 | 33% |
| | VIII | 0.45 | 0.50 | 11% | 0.32 | 0.39 | 22% | 0.23 | 0.31 | 36% |
| | IX | 0.33 | 0.49 | 48% | 0.20 | 0.39 | 95% | 0.12 | 0.31 | 158% |

TABLE 7    SIMULATION RESULTS UNDER NON-COUPLED ARQ

| LSP | | P=0.9 | | | P=0.85 | | | P=0.8 | | |
|---|---|---|---|---|---|---|---|---|---|---|
| Scheme | | PNC-NT | PNC-T | Gain | PNC-NT | PNC-T | Gain | PNC-NT | PNC-T | Gain |
| A T O M | II | 0.76 | 0.78 | 3% | 0.66 | 0.69 | 5% | 0.57 | 0.61 | 7% |
| | V | 0.73 | 0.75 | 3% | 0.61 | 0.66 | 7% | 0.51 | 0.57 | 11% |
| | VI | 0.86 | 0.90 | 4% | 0.69 | 0.73 | 6% | 0.54 | 0.60 | 11% |
| | VII | 0.47 | 0.50 | 6% | 0.36 | 0.40 | 11% | 0.27 | 0.33 | 22% |
| | VIII | 0.61 | 0.65 | 7% | 0.46 | 0.52 | 13% | 0.35 | 0.43 | 23% |
| | IX | 0.60 | 0.69 | 15% | 0.43 | 0.55 | 28% | 0.30 | 0.44 | 46% |

*The unit in Tables 6 and 7 is p/t: packets per time slot.*

For any PNC atom that exploits overheard information, to successfully extract its desired packet, the destination must XOR at least a coded packet (that embeds its desired packet) with an overheard packet (that can be used to extract the desired packet). Recall from the discussion in Section IV.A that, without tracking, the destination can extract its desired packet only when both the coded packet and the overheard packet are successfully received in the same round; with tracking, if the destination receives only a coded packet or an overheard packet that cannot be used to extract a desired packet by itself, the destination will store the coded packet or overheard packet for future use. In a future round, the destination may be able to receive an overheard packet or a coded that can be paired with the stored packet for the extraction of a desired packet. In other words, tracking allows overheard packets received from the past or the future to pair with coded packets, and this increases decoding opportunities at the destination node. In particular, this principle of tracking is common to all PNC atoms with overheard information, and therefore stored-packet tracking is applicable to all PNC atoms that exploit overheard information.

*Performance Gains of Stored-Packet Tracking on Various OPNC Atoms*

We now validate the throughput advantage of store-packet tracking in a number of OPNC atoms by simulations. Since the performance of a PNC atom also depends on whether coupled or non-coupled ARQ is adopted, we compare the PNC scheme with tracking (PNC-T) with the conventional PNC scheme without tracking (PNC-NT) under coupled ARQ (in Table 6) and non-coupled ARQ (in Table 7). We conducted simulations for all PNC atoms with overheard information, that is, atoms II, V (cross atom), VI, VII, VIII (star atom) and IX in [2][13]. Due to space lim-it, in this section we only present the simulation results under homogenous LSP (link success probability). Also, here we only focus on the results under high LSPs, because we are interested in scenarios where PNC has better performance than traditional hop-by-hop transmissions (i.e., we want to compare different variations of PNC, but compare PNC versus the traditional scheme) — as discussed in the first paragraph of Section VI.C, PNC is more efficient than traditional hop-by-hop transmission only when LSP is above certain level (which may be different for different atoms).

---

[13] Since each PNC atom has different structure and transmission pattern (some of them are quite involved), complete presentation of all of them would occupy large space and be a diversion. We refer the reader to our paper [2] for details.

Recall that Section VI showed that by optimizing $W$ and $N$, we can almost compensate for all the throughput degradation due to ACK loss and ACK overhead. Hence, for simplicity, in Parts A and B here, we focus on comparing the benchmark throughputs of different schemes, that is, we assume error-free ACK and neglect the ACK transmission time in the simulations.

Tables 6 and 7 present the throughput gains of PNC-T over PNC-NT. As shown, the throughput gains due to stored-packet tracking for the OPNC atoms listed in the tables are all positive. Also, the throughput gain of an OPNC atom increases with the decrease of LSP. This is because tracking can increase decoding opportunities when wireless channel becomes worse (e.g., tracking will not offer any advantage when LSP=1).

*A.2 General Applicability of Store-Packet Tracking to OSNC*

The main difference between PNC and SNC is whether the network coding operation occurs at the physical layer or at a higher layer. Nevertheless, the same decoding operation (XOR) is performed in both for the extraction of the desired packet. For this reason, since stored-packet tracking can help to increase the chance of successful decoding where there is overhearing, it will also be useful in OSNC atoms that exploit overheard information. Thus, stored-packet tracking can also improve throughput in OSNC and is therefore applicable to OSNC.

B. *General Applicability of Non-Coupled ARQ*

In Section V, for the cross atom, we showed that non-coupled ARQ is more efficient than coupled ARQ with a Markov-model analysis. In principle, we can also construct such Markov models for other atoms. As mentioned in Section IV.B, although the cross atom only has two traffic flows, the number of Markov states is already quite large, making the analysis quite complex. Most PNC atoms have more than two traffic flows, which mean they have even more Markov states, making the analysis even more complex. Therefore, here we compare the efficiency of non-coupled and coupled ARQs through massive simulation experiments.

*B.1 General Applicability of Non-Coupled ARQ to OPNC*

Let us take a look at how non-coupled ARQ can be applied to the star atom. Without loss of generality, we focus on flow A–D. When non-coupled ARQ is adopted, if source A successfully delivers a packet $A_i$ to the destination D in the current round, it will start to send the next packet $A_{i+1}$ in the following round regardless of the other flows. When coupled ARQ is adopted, if the source A successfully delivers a packet $A_i$ to the destination D but either flow B–E or flow C–F

TABLE 8  SIMULATION RESULTS WITHOUT STORED-PACKET TRACKING

| LSP | | P=0.9 | | | P=0.85 | | | P=0.8 | | |
|---|---|---|---|---|---|---|---|---|---|---|
| Scheme | | *Coupled* | *Non-Coupled* | *Gain* | *Coupled* | *Non-Coupled* | *Gain* | *Coupled* | *Non-Coupled* | *Gain* |
| **A T O M** | II | 0.67 | 0.76 | 13% | 0.56 | 0.66 | 18% | 0.46 | 0.57 | 24% |
| | V | 0.62 | 0.73 | 18% | 0.50 | 0.61 | 22% | 0.40 | 0.51 | 28% |
| | VI | 0.65 | 0.86 | 32% | 0.48 | 0.69 | 44% | 0.36 | 0.54 | 50% |
| | VII | 0.35 | 0.47 | 34% | 0.25 | 0.36 | 44% | 0.18 | 0.27 | 51% |
| | VIII | 0.45 | 0.61 | 36% | 0.32 | 0.46 | 44% | 0.23 | 0.35 | 52% |
| | IX | 0.33 | 0.60 | 82% | 0.20 | 0.43 | 115% | 0.12 | 0.30 | 150% |

TABLE 9  SIMULATION RESULTS WITH STORED-PACKET TRACKING

| LSP | | P=0.9 | | | P=0.85 | | | P=0.8 | | |
|---|---|---|---|---|---|---|---|---|---|---|
| Scheme | | *Coupled* | *Non-Coupled* | *Gain* | *Coupled* | *Non-Coupled* | *Gain* | *Coupled* | *Non-Coupled* | *Gain* |
| **A T O M** | II | 0.70 | 0.78 | 11% | 0.60 | 0.69 | 16% | 0.51 | 0.61 | 19% |
| | V | 0.66 | 0.75 | 13% | 0.56 | 0.66 | 17% | 0.48 | 0.57 | 19% |
| | VI | 0.69 | 0.90 | 29% | 0.54 | 0.73 | 32% | 0.43 | 0.60 | 37% |
| | VII | 0.38 | 0.50 | 30% | 0.30 | 0.40 | 33% | 0.24 | 0.33 | 38% |
| | VIII | 0.50 | 0.65 | 30% | 0.39 | 0.52 | 35% | 0.31 | 0.43 | 40% |
| | IX | 0.49 | 0.69 | 40% | 0.39 | 0.55 | 41% | 0.31 | 0.44 | 42% |

*The unit of throughput in Tables 8 and 9 is **p/t**: packets per time slot.*

fails delivering its current packet $B_i$ or $C_i$ in the same round, it will still send the delivered packet $A_i$ in the following round. A source will not transmit the next packet indexed by $i+1$ until all sources have successfully delivered their packets indexed by $i$.

We have tested different atoms under different LSPs and the simulations are conducted without tracking (in Table 8) and with tracking (in Table 9). We found that for all PNC atoms with overheard information, non-coupled ARQ is more efficient than coupled ARQ whether packet tracking is adopted or not. As shown in Tables 8 and 9, the throughput gains range from 11% to 150%.

*B.2 General Applicability of Non-Coupled ARQ to OSNC*

In Section VII.A.2, we pointed out that an SNC atom does the same decoding operation as its corresponding PNC atom. In particular, whether an atom with a particular topological structure utilizes PNC or SNC, a specific destination (e.g., destination C in the cross atom) decodes the same coded packet (e.g., $A \oplus B$ in PNC or $A \oplus B$ SNC) from the relay with the same overheard packet (e.g., $B$ in PNC or $B$ in SNC). That means the packet delivery of a specific traffic flow (e.g. traffic flow A-C) depends on the overheard information from the same other traffic flow (e.g., traffic flow B-D). In other words, the dependencies among different traffic flows in an SNC atom are the same as in its corresponding PNC atom. For example, in both PNC and SNC, flow A-C is dependent on flow B-D in that to extract its desired packet $A$, destination C needs the overheard packet from source $B$.

Intuitively, since the dependencies among different traffic flows are the same, for any atom, as long as the atom utilizing PNC prefers non-coupled ARQ, the corresponding SNC atom would also prefer non-coupled ARQ. Furthermore, for a specific round, the possible throughput of adopting non-coupled ARQ relative to the possible throughput of adopting coupled ARQ are the

TABLE 10  THROUGHPUTS COMPARISON FOR PNC STAR ATOM

|  | *P=0.95* | *P=0.9* | *P=0.85* | *P=0.8* | *P=0.75* |
|---|---|---|---|---|---|
| **Benchmark** (*p/t*) | 0.80 | 0.65 | 0.52 | 0.43 | 0.35 |
| **PNC-O (N=3 W=130)** (*p/t*) | 0.77 | 0.63 | 0.51 | 0.42 | 0.33 |
| *Degradation1* | 3.4% | 2.9% | 2.1% | 2.4% | 2.6% |
| **PNC (N=1 W=1)** (*p/t*) | 0.77 | 0.63 | 0.51 | 0.42 | 0.33 |
| *Degradation2* | 14% | 23% | 33% | 42% | 56% |

*p/t: packets per time slot.*

same whether PNC or SNC is exploited. Hence, the throughput gains of non-coupled ARQ over coupled ARQ for PNC atoms (in Tables 8 and 9) are also applicable to SNC atoms.

*C. General Applicability of optimizing W and N*

The performance of atoms other than the cross atom will also benefit from the optimization of transmission window size ($W$) and ACK frequency ($1/N$) in their ARQ. Take the PNC star atom for example. By the method proposed in Section VI.C, we found that a good set of $W$ and $N$ for the star atom are $W=130$ and $N=3$ when $K=30$ bytes and $D=600$ bytes (note: this set of ($W$, $N$) is different from the set for the cross atom). These $W$ and $N$ can ensure that the worst throughput degradation compared with the upper bound benchmark (i.e., the non-coupled simulation results in Table 9) is less than 4% for different $p$ (See Table 10). But if we just set $W=1$ and $N=1$, the throughput degradations range from 14% to 56% for the same $p$ range. Hence, proper optimizing $W$ and $N$ is important in the ARQ design of the PNC star atom.

*Generality of Optimizing W and N*

In general, optimizing transmission window size and ACK frequency is applicable to all network coded systems that make use of ARQ to ensure reliability; i.e., it is not just limited to our study of PNC here. By optimizing the transmission window size $W$ and the ACK frequency $1/N$, we can reduce the throughput degradation due to lost ACK and ACK overhead. In a practical network-coded ARQ system, lost ACK and ACK overhead are part of the system, whether we are talking about a system operated with PNC or SNC. Thus, optimizing $W$ and $N$ can potentially minimize this throughput degradation in all network coded systems in which ACK is error-prone and ACK overhead cannot be ignored.

To our best our knowledge, there has been no past network-coded work that considered the issue of optimizing transmission window size and ACK frequency in a rigorous manner (e.g., [4] and [5]). Hence, going forward, it will be interesting to investigate how to set $W$ and $N$ in different network coded systems taking into consideration the effect of error-prone ACK and ACK overhead.

## VIII. CONCLUSION

This paper has investigated ARQ (Automatic Repeat request) designs for PNC (Physical-layer Network Coding) systems. Specially, we focus on PNC systems that exploit overheard information.

Our investigation indicates that for PNC ARQ: 1) Storing both coded and overheard packets and their proper tracking can increase the chance of packet extraction, hence throughput. 2) Between coupled ARQ and non-coupled ARQ, two generic variations of PNC ARQ, non-coupled ARQ is more efficient than coupled ARQ. 3) Proper optimization of the transmission window size and ACK frequency can nearly compensate for all the throughput loss due to lost ACK and ACK overhead (the resulting throughput degradation is within 4% of a theoretical throughput upper bound).[14]

As a first investigation of ARQ for PNC building blocks that exploit overheard information, this paper has only focused on end-to-end ARQ. Going forward, it will also be interesting to investigate link-by-link ARQ and compare the relative merits of these two ARQ approaches for PNC systems.

---

[14] The three ARQ principles, stored-packet tracking, non-coupled ARQ, and window-size and ACK frequency optimization, are also applicable to SNC. In other words, they can also improve performance in SNC systems.